\documentstyle[12pt,epsf]{article}
\def\beq{\begin{equation}}
\def\eeq{\end{equation}}
\def\barr{\begin{eqnarray}}
\def\earr{\end{eqnarray}}

\def\b{\bigskip}

\begin{document}

\title{Edge Asymptotics of Planar Electron
Densities\footnote{UCONN-93-9;hep-th/9311141}}
\author{Gerald V. Dunne \\
Department of Physics\\
University of Connecticut\\2152 Hillside Road\\
Storrs, CT 06269 USA\\
   \\
dunne@hep.phys.uconn.edu \\}
\date{November 10, 1993}

\maketitle
\begin{abstract}The $N\to\infty$ limit of the edges of finite planar electron
densities is
discussed for higher Landau levels. For full filling, the particle number is
correlated with the magnetic flux, and hence with the boundary location, making
the $N\to\infty$ limit more subtle at the edges than in the bulk. In the
$n^{\rm th}$ Landau level, the density exhibits $n$ distinct steps at the edge,
in both circular and rectangular samples. The boundary characteristics for
individual Landau levels, and for successively filled Landau levels, are
computed in an asymptotic expansion.
\end{abstract}
\section{Introduction}
\label{sec-intro}
Recent work \cite{halp,wen,sto,zee,cdtz,cap,iso} has illustrated the
fundamental importance of the {\it edge} properties of planar incompressible
quantum fluids in the understanding of the quantum Hall effect \cite{pra}. The
physical incompressibility of the quantum Hall samples correlates boundary and
bulk degrees of freedom in an interesting manner. The boundary values of the
bulk particle densities behave as $1+1$ dimensional chiral Kac-Moody currents,
the algebraic properties of which characterize the edge excitations
\cite{wen,sto,cdtz}. Furthermore, the incompressibility may be fruitfully
understood via a quantum deformation of the algebra of area-preserving
diffeomorphisms \cite{cap,iso,mar}. Most of this work has concentrated on the
physics of the lowest Landau level, but higher Landau levels are also important
for a complete description of the quantum Hall effect. Here, I consider the
exact large $N$ asymptotics of the edge properties of incompressible quantum
fluids in higher Landau levels.

Consider the nonrelativistic Pauli hamiltonian operator for spin-polarized
(i.e. scalar) fermions confined to the two-dimensional plane, in the presence
of a uniform magnetic field (of strength B) perpendicular to the plane:
\beq
H={1\over 2m}\left(\vec{p} -{e\over c} \vec{A}\right)^2 -{e\hbar B\over 2mc}
\label{ham}
\eeq
It is well known that, in the plane, the spectrum of this hamiltonian consists
of an infinite discrete set of equally spaced energy levels (called "Landau
levels"), each of which is infinitely degenerate \cite{land}. The infinite
degeneracy of each level is directly related to the infinite area of the plane,
and for a sample of finite area the degeneracy $N$ is of the order of the net
magnetic flux $\phi$ through the region (measured in fundamental flux units $h
c/e$):
\beq
N=\phi + \mu
\label{deg}
\eeq
where $\mu=O(1)$. In the plane, the eigenstates of the energy operator
(\ref{ham}) may be written as $\psi_n^j$, where $n=0,1,2,\dots$ is the Landau
level index (corresponding to the energy eigenvalue $E_n=n \hbar \omega_c$
where $\omega_c=eB/mc$ is the cyclotron frequency) and $j$ labels the
degenerate states within each Landau level. In the $n^{th}$ Landau level, the
expectation value of the density operator is given by the function
\beq
\rho_n(\vec{x}) = \sum_{j} |\psi_n^j(\vec{x})|^2
\label{den}
\eeq
For a {\it fully filled} Landau level, the occupation number equals the
degeneracy, so the summation in (\ref{den}) is always over $N$ values of $j$.
At some fixed point $\vec{x}$ within the bulk of the sample, the thermodynamic
limit, $N\to\infty$, simply involves extending this finite sum to an infinite
sum. But at the {\it edge} of the sample, each individual density
$|\psi_n^j(\vec{x})|^2$ in the sum is itself evaluated at a distance scale
$|\vec{x}|=R$ which is correlated with the particle number as $R\sim\sqrt{\phi}
\sim\sqrt{N}$ because of (\ref{deg}) and the condition of full filling. Thus
the thermodynamic limit is much more subtle at the sample boundaries. The
analysis of these limits is the subject of this paper.

The precise form of the eigenstates $\psi_n^j$ depends on the gauge chosen for
the vector potential $\vec{A}$. To describe a finite sample, one should choose
the gauge with the geometry of the sample boundary in mind. It is convenient to
choose the gauge so that, on the boundary, the normal component of the vector
potential $\vec{A}$ vanishes. Thus, for circular samples one uses the
"symmetric" gauge, $\vec{A}=-{B\over 2} (y,-x)$, while for rectangular samples
(actually for rectangular strips with upper and lower sides identified) one
uses the "Landau" gauge, $\vec{A}=B (0,x)$. Owing to the degeneracy of the
Landau levels, a change in the gauge involves a change in the nature of the sum
over $j$ in (\ref{den}), and so the functional form of $\rho_n$ is gauge
dependent. We shall see that it is considerably easier to analyze the
asymptotics of $\rho_n$ in the symmetric gauge than in the Landau gauge.

\section{Circular Geometry}
\label{sec-circle}

In the symmetric gauge it is useful to work with complex coordinates $z\equiv
x+iy$, and the degenerate states are labelled by an (integer) angular momentum
index $j$:
\beq
\psi_n^j ={1\over \sqrt{\pi}}\left({B\over 2}\right)^{{j+1\over2}}
\sqrt{{n!\over(n+j)!}}~z^j~L_n^j\left({Br^2 \over 2}\right)
exp\left({-{Br^2\over 4}}\right)
\label{lag}
\eeq
In (\ref{lag}), the Landau level index $n$ runs over all nonnegative integers
$n=0,1,2,\dots$, while the angular momentum index takes values
$j=-n,-n+1,-n+2,\dots$ within the $n^{th}$ energy level. The $L_n^j$ are the
generalized Laguerre polynomials \cite{bat} and the eigenstates $\psi_n^j$ are
orthonormal in the plane. The individual densities $|\psi_n^j(\vec{x})|^2$ are
radially symmetric functions of Gaussian form, peaked at a radius
$r\simeq\sqrt{j+n}$, which increases with $j$. So for a finite disc
\footnote{In a similar way, one can of course treat an annulus as the {\it
difference} of two discs.} of $N$ electrons, all in the $n^{th}$ Landau level,
the density function is \footnote{The overall normalization has been chosen so
that $\int d\xi \rho_n=N$, the degeneracy number.}
\beq
\rho_n(\xi, N) = n!~e^{-\xi} \sum_{j=-n}^{N-n-1} {\xi^j\over
(n+j)!}\left(L_n^j(\xi)\right)^2
\label{lden}
\eeq
where the natural dimensionless variable is $\xi={B\over 2}r^2$, in terms of
which the boundary $r=R$ is given by $\xi={B\over 2}R^2\equiv\phi$. Thus,
analysis of the boundary density in the thermodynamic limit involves the
$N\to\infty$ behaviour of $\rho_n(\xi, N)$ for $\xi=\phi\simeq N$.

\begin{figure}
    \epsffile{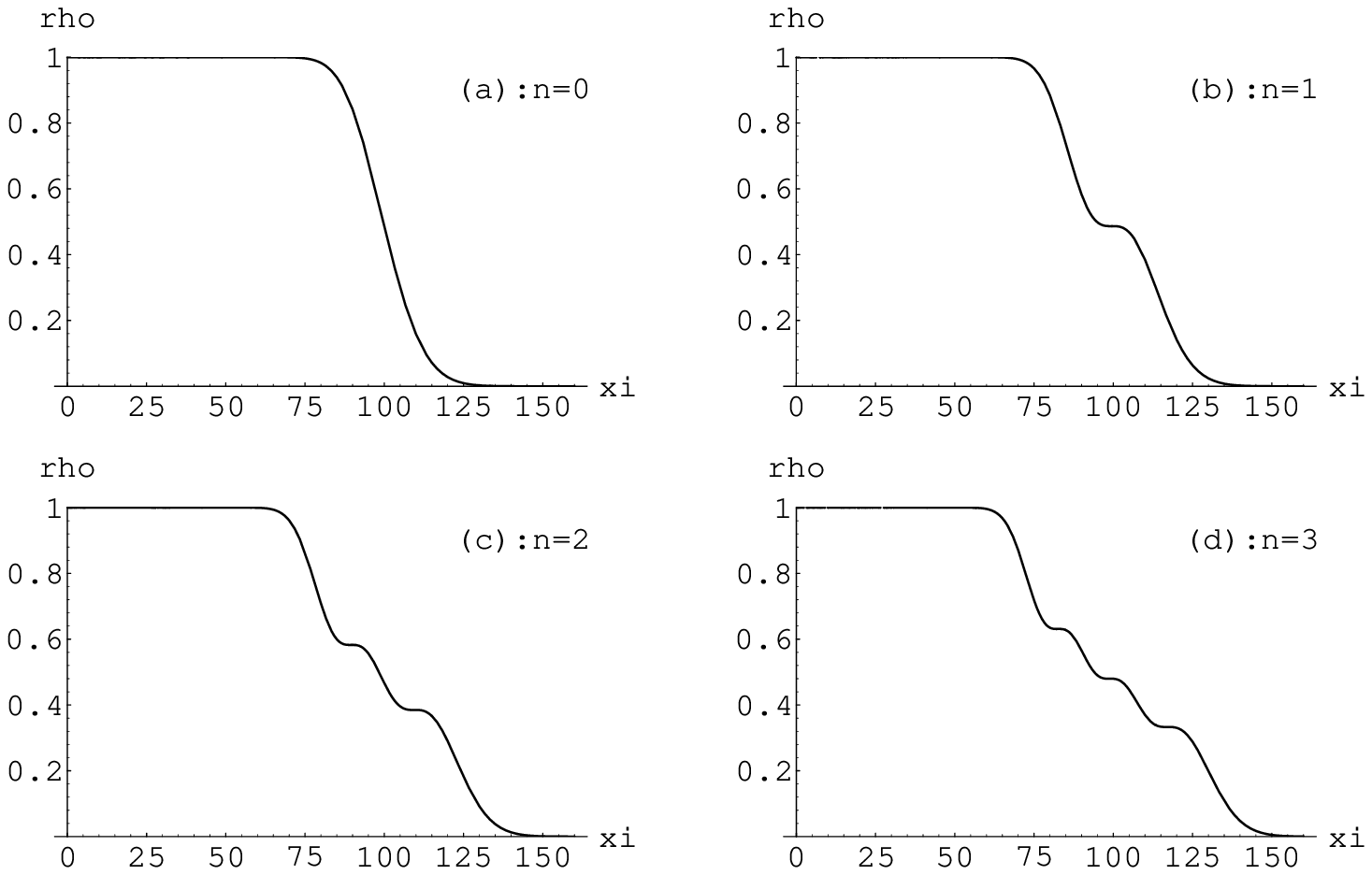}
    \caption{The symmetric gauge density functions (\protect{\ref{lden}})
plotted for $N=100$ electrons for the first four Landau levels, $n=0,1,2,3$.
Notice the uniform bulk density equal to $1$ and the boundary fall-off at
$\xi\simeq N$. For the $n^{th}$ Landau level there are $n$ distinct steps at
the boundary. The exact locations of these steps are given (see equation
(\protect{\ref{der}})) by the zeros of $L_n^{N-n-1}$ and $L_n^{N-n}$.}
  \label{symmetricplot}
\end{figure}

Consider first of all the density function for the lowest Landau level,
\barr
\rho_0(\xi, N)&=&e^{-\xi} \sum_{j=0}^{N-1} {\xi^j\over j!}\nonumber\\
 &=&e^{-\xi} e_{N-1}(\xi),
\label{lll}
\earr
where $e_{N-1}(\xi)$ is the truncated exponential function \cite{bat}. This
lowest Landau level density function is plotted in Figure (\ref{symmetricplot}
a) for $N=100$. We see that $\rho_0(\xi, N)$ does indeed represent a finite
droplet of uniform bulk density (equal to 1 in the normalization chosen in
equation (\ref{lden})), dropping rapidly to zero at the 'edge' where $\xi
\simeq N$. For fixed $\xi$ {\it within the bulk} it is clear from (\ref{lll})
that $\rho_0(\xi, N)\to 1$ as $N\to\infty$. Indeed, since the bulk density is
asymptotically uniform (see below), we can evaluate this uniform value deep in
the bulk ({\it i.e.} at $\xi =0$) and (\ref{lll}) immediately gives $\rho_0(0,
N)=1$.

To study $\rho_0(\xi, N)$ in the vicinity of the boundary ({\it i.e.} evaluated
at $\xi=\phi=N-\mu$), we use the following integral representation \cite{bat}
for the truncated exponential function in terms of the incomplete Gamma
function $\Gamma(N, \xi)$
\beq
e_{N-1}(\xi)={\Gamma(N, \xi)\over \Gamma(N)}e^{\xi}
\label{trunc}
\eeq
where
\beq
\Gamma(N, \xi) \equiv \int_{\xi}^{\infty} e^{-t} t^{N-1} dt
\label{gamma}
\eeq
One can then use Laplace's method \cite{cmb} to show that the large $N$
behaviour near the edge is
\beq
\rho_0(N-\mu, N) = {1\over 2} +{(\mu-{1\over 3})\over \sqrt{2\pi N}} +
O({1\over N})
\label{laplace}
\eeq
giving a leading order edge density equal to half the bulk density, with
subleading corrections that go as $O({1\over \sqrt{N}})$ in the variable $\xi$.
The $\xi$ derviative of the lowest Landau level density is
\beq
{d\over d\xi}\rho_0=-{\xi^{N-{1}}\over (N-1)!}~e^{-\xi}
\label{lllslope}
\eeq
Note that the summation over $j$ has disappeared, due to cancellations between
successive terms, a fact that makes the $N\to\infty$ limit of the derivative
much easier to consider. For large $N$, the derivative is nonzero only in the
vicinity of the boundary. In fact, for large $N$, all derivatives of $\rho_0$
vanish in the bulk, explaining the uniform bulk density. We can estimate the
derivative ${d\over d\xi}\rho_0$ at the boundary using the fact
\beq
{\xi^N\over N!}~e^{-\xi} = {1\over \sqrt{2\pi N}} e^{-{(\xi-N)^2\over 2N}}
\left(1-{1\over 12N}+\dots \right)\left(1+{1\over 3}{(\xi-N)^3 \over
N^2}+\dots\right)
\label{gauss}
\eeq
Therefore,
\beq
{d\over d\xi}\rho_0 (N-\mu, N) = - {1\over \sqrt{2\pi N}}
e^{-\mu^2/2N}\left(1+{\mu-1/12 \over N}+\dots \right)
\eeq
However, since the sample itself grows in size as $N$ increases
($R\sim\sqrt{N}$), it is more meaningful to consider the derivative in terms of
the rescaled variable $\zeta=\xi/N$, for which the boundary is at
$\zeta=(1-{\mu\over N})$. In terms of this rescaled coordinate,
\beq
{d\over d \zeta} \rho_0 |_{\zeta=1-\mu/N}=-\sqrt{{N\over
2\pi}}e^{-N(\mu/N)^2/2}\left(1+O({1\over N})\right)
\label{zetaderiv}
\eeq
which shows that the edge derivative becomes large and negative as $-\sqrt{N}$;
and that the derivative is a delta function of width ${1\over {N}}$
concentrated at the boundary, as is appropriate for the derivative of a
step-function density profile. In terms of the actual radial coordinate
$r=\sqrt{{2\over B}\xi}$, the slope at the boundary is asymptotically constant
\beq
{d\over dr}\rho_0 \big{|}_{\rm boundary} =-\sqrt{B\over\pi}\left(1+O({1\over
N})\right)
\label{rderiv}
\eeq

These results for the lowest Landau level density are reassuring, but not
particularly surprising. The $N\to\infty$ limit is greatly facilitated by the
relation of $\rho_0(\xi, N)$ to the incomplete gamma function. Such a direct
relation is not available for the higher Landau levels due to the appearance of
the Laguerre polynomial factors in (\ref{lden}). However, the asymptotic edge
behaviour is significantly more interesting for these higher Landau levels. In
Figure (\ref{symmetricplot}), the densities $\rho_n$ are plotted for
$n=0,1,2,3$ with $N=100$. The pattern is striking - each density function has
uniform bulk value equal to unity, and each falls off at an edge located at
$\xi \simeq N$, but the $n\geq 1$ densities exhibit a distinct step-like
pattern. This is a general feature of $\rho_n(\xi, N)$ - there are exactly $n$
such steps in the vicinity of the boundary. However, if one looks more closely
by magnifying the scale of $\rho_n$ considerably, these "steps" are in fact
pairs of a local minimum and a local maximum. The locations of these local
minima and maxima may be identified as a result of the following remarkable
identity which generalizes the formula (\ref{lllslope}):
\beq
{d\over d\xi} \rho_n (\xi, N) = - n! e^{-\xi}
{\xi^{N-n-1}\over(N-1)!}L_n^{N-n-1}(\xi)~L_n^{N-n}(\xi)
\label{der}
\eeq
Note that, as in (\ref{lllslope}), taking the derivative of $\rho_n$ enables
the removal of the summation over the angular momentum index: this is due to
intricate cancellations between successive terms in the sum, using the
recurrence and differential-difference relations of the Laguerre polynomials.
{}From (\ref{der}) we see that the 'troughs' and 'peaks' of each step are
located at the zeros of $L_n^{N-n+1}(\xi)$ and $L_n^{N-n}(\xi)$. Each Laguerre
factor in (\ref{der}) is a polynomial of degree $n$ with $n$ real zeros, and,
furthermore, their zeros are interwoven (and for a given $n$, the $k^{th}$ zero
of $L_n^j$ increases with $j$ \cite{bat}), explaining the pairing of local
maxima and minima which gives the impression of steps.

\begin{figure}
\epsffile{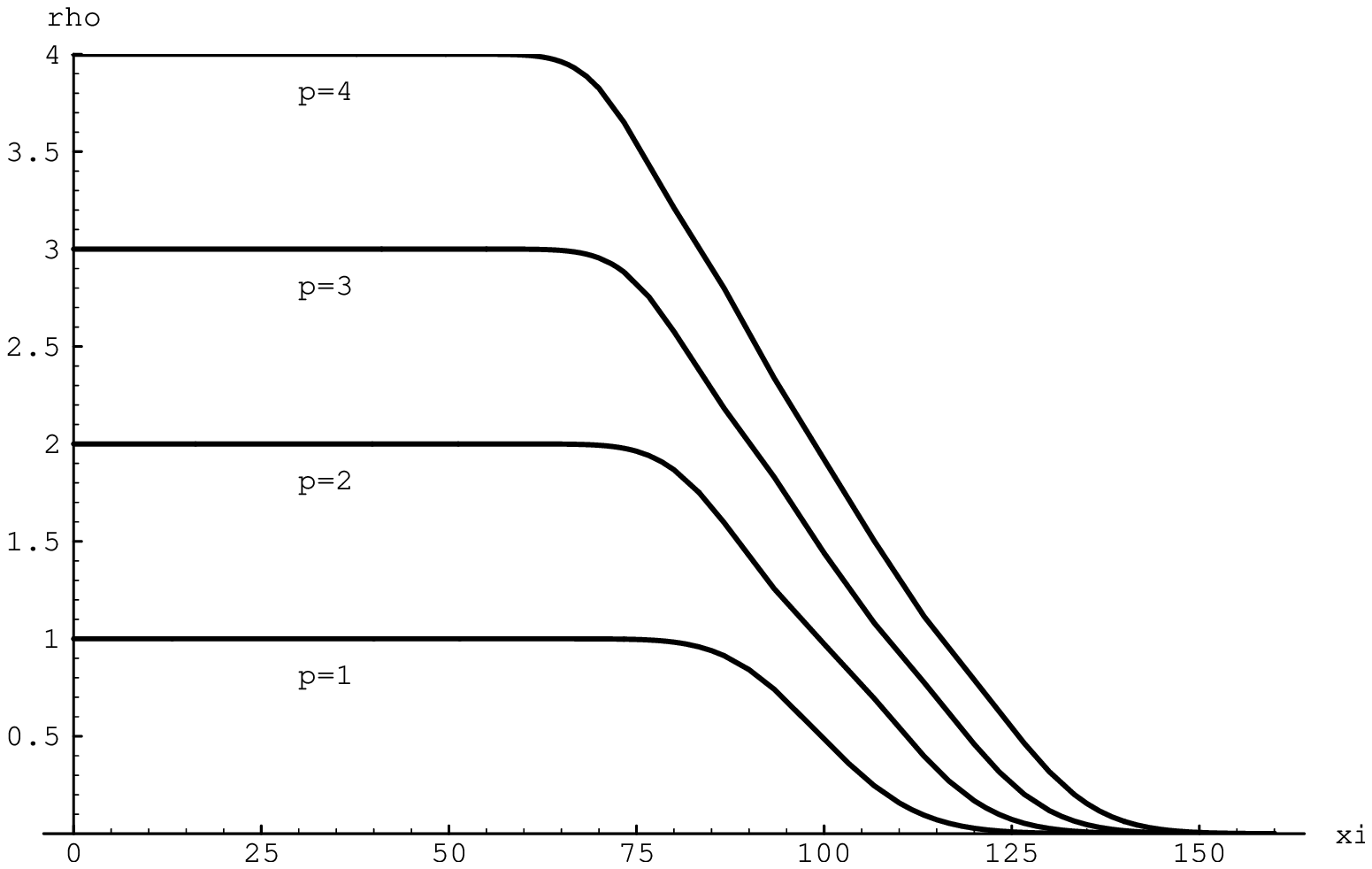}
\caption{The symmetric gauge density functions $\rho_{\rm total}^{(p)}(\xi, N)$
for $p$ fully filled Landau levels, plotted for $N=100$ electrons. Notice the
uniform bulk density equal to $p$, the boundary density equal to ${p\over 2}$,
and the absence of the boundary steps seen in the individual higher Landau
level densities in Figure (\protect{\ref{symmetricplot}}).}
\label{fullplot}
\end{figure}

\begin{figure}
\epsffile{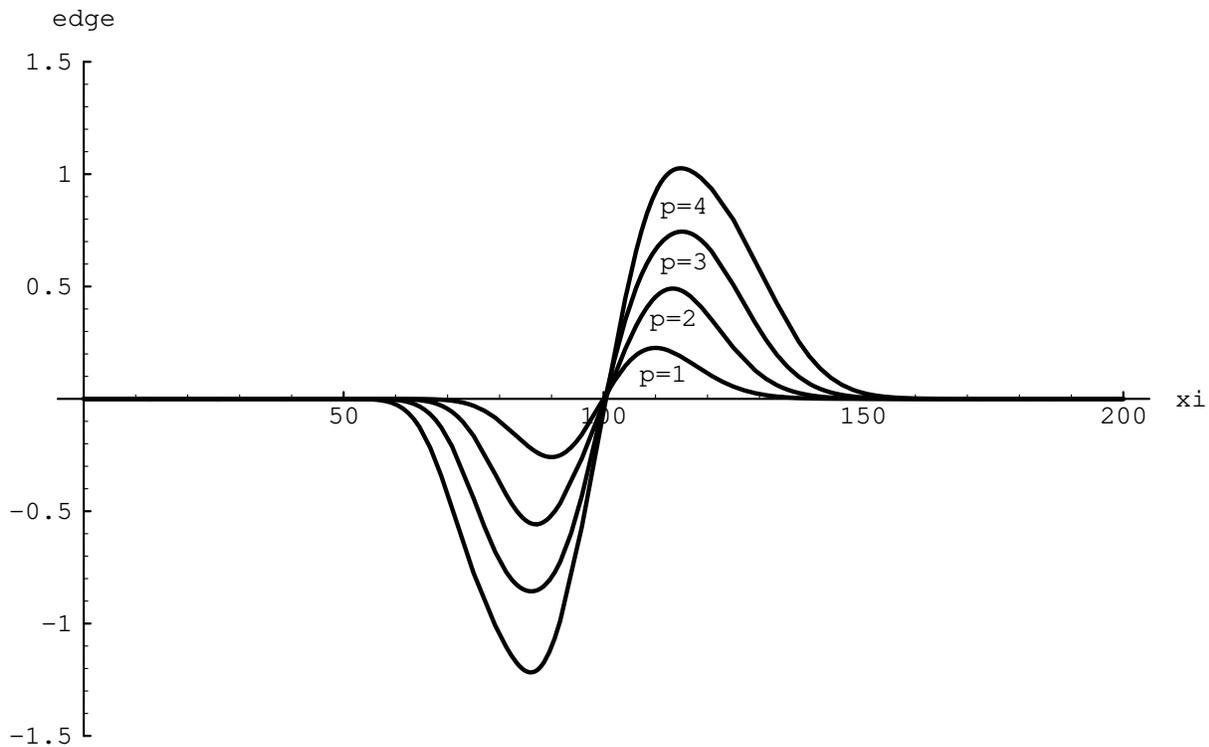}
\caption{The $N=100$ edge contribution by which the symmetric gauge density for
$p$ fully filled Landau levels differs from $p$ times the density for the
lowest Landau level - see (\protect{\ref{pdensity}},\protect{\ref{pex}}).
Notice that the edge contribution is centered at $\xi\simeq 100$ and that for
different $p$ it has the same functional form.}
\label{edgeplot}
\end{figure}

Perhaps even more remarkable than this step-like behaviour is the fact that
when the density functions for successive Landau levels are summed, the
boundary again becomes "smooth" (i.e. the steps disappear). This is directly
relevant for applications to the quantum Hall effect \cite{pra}, where one is
interested, for example, in the density for full filling of the first $p$
Landau levels:
\beq
\rho_{\rm total}^{(p)} (\xi, N) \equiv \sum_{n=0}^{p-1} \rho_n (\xi, N)
\label{full}
\eeq
These fully filled densities are plotted in Figure (\ref{fullplot}) for
$p=1,2,3,4$ and for $N=100$. Notice that the form of these density functions is
once again like the lowest Landau level density function $\rho_0$ plotted in
Figure (\ref{symmetricplot} a). To explain this miraculous cancellation of the
boundary steps, note first of all the following identity \footnote{This is a
nontrivial identity which relies on the recurrence relation properties of the
Laguerre polynomials.} which relates the density in the $n^{th}$ level to the
density in the $(n-1)^{th}$ level:
\beq
\rho_n (\xi, N) =\rho_{n-1} (\xi, N) - (n-1)! e^{-\xi} {\xi^{N-n} \over (N-1)!}
L_n^{N-n} (\xi) L_{n-1}^{N-n} (\xi)
\label{induction}
\eeq
The second term on the right hand side is concentrated at the boundary, since
it is a sum of terms like (\ref{gauss}). Using this identity, it follows that
\beq
\rho_{\rm total}^{(p)} (\xi, N) = p~\rho_0 (\xi, N) - {e^{-\xi}\over
(N-1)!}\sum_{n=1}^{p-1} (p-n)(n-1)!\xi^{N-n} L_n^{N-n} (\xi) L_{n-1}^{N-n}
(\xi)
\label{pdensity}
\eeq
Once again, the sum on the right hand side is concentrated at the boundary,
while the bulk value is just $p$ (i.e., $p$ times the bulk value of $\rho_0$).
For low values of $p$ it is easy to perform the finite sums in
(\ref{pdensity}), yielding
\barr
\rho_{\rm total}^{(1)}(\xi, N)&=&\rho_0(\xi, N) \nonumber\\
\rho_{\rm total}^{(2)}(\xi, N)&=&2 \rho_0 (\xi, N) + e^{-\xi} {\xi^{N-1}\over
(N-1)!} \left(\xi-N\right)\nonumber\\
\rho_{\rm total}^{(3)}(\xi, N)&=&3 \rho_0 (\xi, N) + e^{-\xi} {\xi^{N-2}\over
2(N-1)!} \left(\xi^3-(3N-5)\xi^2+N(3N-7)\xi-N(N-1)^2\right)\nonumber\\
\rho_{\rm total}^{(4)}(\xi, N)&=&4 \rho_0 (\xi, N) + e^{-\xi} {\xi^{N-3}\over
6(N-1)!} \left(\xi^5-(5N-8)\xi^4+(10N^2-30N+26)\xi^3 \right.\nonumber\\
& & \left.
\hspace{1in}-N(10N^2-42N+50)\xi^2+N(5N^3-26N^2+37N-16)\xi\right.\nonumber\\
& & \left. \hspace{1in}-N(N-1)^2(N-2)^2\right)
\label{pex}
\earr
The edge contributions are all concentrated at the boundary, and despite the
seemingly complicated expressions (\ref{pex}), they all have a common
functional form - see Figure (\ref{edgeplot}).

Thus, the leading behaviour of the density of $p$ fully filled Landau levels is
identical to $p$ times the density of the fully filled lowest Landau level,
with subleading corrections $\sim O(1/\sqrt{N})$ concentrated at the boundary.
For full filling of the first few Landau levels, this subleading edge behaviour
can be determined from (\ref{laplace}) and (\ref{pex}):
\barr
\rho_{\rm total}^{(1)}(N-\mu, N)&=&{1\over 2}+{\mu-{1\over 3} \over\sqrt{2\pi
N}}+O({1\over N})\nonumber\\
\rho_{\rm total}^{(2)}(N-\mu, N)&=&1+{\mu-{2\over 3} \over\sqrt{2\pi
N}}+O({1\over N})\nonumber\\
\rho_{\rm total}^{(3)}(N-\mu, N)&=&{3\over 2} + {{3\over 2}\mu-{3\over 2}
\over\sqrt{2\pi N}}+O({1\over N})\nonumber\\
\rho_{\rm total}^{(4)}(N-\mu, N)&=&2 + {{3\over 2}\mu-2 \over\sqrt{2\pi
N}}+O({1\over N})
\label{boundary}
\earr
The general formula for this subleading behaviour, for {\it any} $p$, is
\barr
\rho_{\rm total}^{(p)}(N-\mu, N)&=&{p\over 2}+{2\Gamma\left(\left[{p+1\over
2}\right] +{1\over 2}\right) \over \Gamma\left({1\over 2}\right)
\Gamma\left(\left[{p+1\over 2}\right]\right)} \left({\mu-{p\over 3}
\over\sqrt{2\pi N}}\right)+O({1\over N})
\label{gen}
\earr
where $[{p+1\over 2}]$ denotes the integer part of ${p+1\over 2}$. Notice that
the coefficient of ${\mu\over \sqrt{2 \pi N}}$ for $p$ odd is equal to that for
$p+1$.

The derivative of $\rho_{\rm total}^{(p)}$ is localized at the boundary, since
\beq
{d\over d\xi}\rho_{\rm total}^{(p)} =- {e^{-\xi} \over (N-1)!} \sum_{n=0}^{p-1}
n! \xi^{N-n-1} L_{n}^{N-n-1}(\xi) L_n^{N-n} (\xi)
\label{pslope}
\eeq
For low values of $p$, this derivative is:
\barr
{d\over d\xi} \rho_{\rm total}^{(1)}&=&- e^{-\xi} {\xi^{N-1}\over
(N-1)!}\nonumber\\
{d\over d\xi} \rho_{\rm total}^{(2)}&=&- e^{-\xi} {\xi^{N-2}\over
(N-1)!}\left(\xi^2-2(N-1)\xi +N(N-1)\right)\nonumber\\
{d\over d\xi} \rho_{\rm total}^{(3)}&=&- e^{-\xi} {\xi^{N-3}\over 2(N-1)!}
\left(\xi^4-4(N-1)\xi^3+6(N-1)^2\xi^2 -4N(N-1)(N-2)\xi\right.\nonumber\\
& & \hspace{3in}\left.+N(N-1)^2(N-2)\right)\nonumber\\
{d\over d\xi} \rho_{\rm total}^{(4)}&=&- e^{-\xi} {\xi^{N-4}\over 6(N-1)!}
\left(\xi^6-6(N-1)\xi^5-3(N-1)(5N-6)\xi^4\right.\nonumber\\
& &\hspace{.5in}\left.+4(N-1)(N-2)(5N-3)\xi^3-
3N(N-1)(N-2)(5N-11)\xi^2\right.\nonumber\\
& &\hspace{.5in}\left.+6N(N-1)^2(N-2)(N-3)\xi-N(N-1)^2(N-2)^2(N-3)\right)
\label{deriv}
\earr
In terms of the radial coordinate, the asymptotic boundary slopes are
\barr
{d\over dr} \rho_{\rm total}^{(1)}|^{\rm boundary}&=&-\sqrt{B\over
\pi}\left(1+O({1\over N})\right) \nonumber\\
{d\over dr} \rho_{\rm total}^{(2)}|^{\rm boundary}&=&-\sqrt{B\over
\pi}\left(1+O({1\over N})\right) \nonumber\\
{d\over dr} \rho_{\rm total}^{(3)}|^{\rm boundary}&=&-{3\over 2}\sqrt{B\over
\pi}\left(1+O({1\over N})\right) \nonumber\\
{d\over dr} \rho_{\rm total}^{(4)}|^{\rm boundary}&=&-{3\over 2}\sqrt{B\over
\pi}\left(1+O({1\over N})\right)
\label{edgederiv}
\earr
The general formula for the leading-order boundary slope is
\beq
{d\over dr} \rho_{\rm total}^{(p)}|^{\rm boundary}=
-\sqrt{{B\over \pi}}{2\Gamma\left(\left[{p+1\over 2}\right] +{1\over 2}\right)
\over \Gamma\left({1\over 2}\right) \Gamma\left(\left[{p+1\over
2}\right]\right)} \left( 1+ O\left({1\over N}\right)\right)
\label{ppslope}
\eeq
where, as before, $[{p+1\over 2}]$ denotes the integer part of ${p+1\over 2}$.
Notice that for $p$ odd the asymptotic boundary slope is equal to that for
$p+1$. Note also that the leading dependence of the slope agrees with the
$O(\mu/\sqrt{N})$ dependence of $\rho_{\rm total}^{(p)}(N-\mu, N)$ given in
(\ref{boundary},\ref{gen}), as it must.

These results rely on such specific properties of the Laguerre polynomials,
that it is not immediately clear whether these strange edge phenomena are some
sort of side-effect of the symmetric gauge or of the circular nature of the
samples. It is therefore instructive to consider also the boundary phenomena
for the rectangular samples using the Landau gauge. Notice that the Landau and
symmetric gauges are related by the gauge transformation
$\vec{A}^{Landau}=\vec{A}^{symmetric}+\vec{\nabla}\lambda$ with gauge function
$\lambda={B\over 2}xy$. Normally, in the absence of degeneracies, different
gauges lead trivially to identical densities since the nondegenerate
wavefunctions are simply related by a phase factor $e^{i\lambda}$. However, due
to the degeneracy of the Landau levels, the gauge invariance only implies that
$e^{i\lambda}\psi_n^{j,{}symmetric}$ is some linear combination of the
$\psi_n^{k,{}Landau}$ with the same energy ({\it i.e.} with the same Landau
level index $n$). The particular form of this linear combination is gauge
dependent, and so the conversion of the density from one gauge to another is
not such a straightforward matter.

\section{Rectangular Geometry}
\label{sec-square}

In the Landau gauge, the eigenstates of the Pauli energy operator (\ref{ham})
are
\beq
\psi_n^k = e^{iky}\left({B\over \pi}\right)^{1/4}{1 \over \sqrt{2^n n!}} H_n
\left(\sqrt{B}(x-{k\over B})\right) exp\left(-{B\over 2}\left(x-{k\over
B}\right)^2\right)
\label{her}
\eeq
where, as in (\ref{lag}), the Landau level label $n=0,1,2,\dots $, but now the
degenerate states are labelled by a continuous momentum index $k$. This
continuous momentum label may be discretized by compactifying in the $y$
direction. That is, if we consider the infinite strip $-\infty<x<+\infty$,
$0\leq y\leq L$ rather than the entire plane, then with periodic or
anti-periodic boundary conditions in the $y$ direction, the momentum index $k$
takes values $k=({2\pi\over L}) j$. Here, $j$ takes integer values for periodic
boundary conditions, and half-odd-integer values for anti-periodic boundary
conditions.   The $H_n$ are the Hermite polynomials \cite{bat}, and the
wavefunctions $\psi_n^k$ are orthogonal in the infinite strip.

The individual densities $|\psi_n^k|^2$ are independent of the $y$ coordinate
(just as in the symmetric gauge the individual densities are independent of the
angular coordinate), and have a Gaussian form in the $x$ direction (analogous
to the radial Gaussian form of the densities in the symmetric gauge) peaked at
$x\simeq {k\over B}$. Thus, for a rectangular strip of finite extent in the $x$
direction,\footnote{Note that this means that the topological equivalent is an
annulus rather than a disc, but an annular density may be simply treated as the
"difference" of two discs.} there are corresponding upper and lower bounds on
the allowed values of the discrete momentum.\footnote{This is Landau's original
argument for estimating the degeneracy in terms of the magnetic flux through
the finite sample \cite{land}.} For convenience of notation, we choose the
sample to be a {\it square}: $-{L\over 2}\leq x\leq+{L\over 2}$, $0\leq  y\leq
L$, and use the rescaled $x$-coordinate, $\xi \equiv {BL\over 2\pi} x =
{\phi\over L} x$, in terms of which the boundaries lie at $\xi=\pm {\phi\over
2}$.

Then the density for the fully filled $n^{th}$ Landau level is
\beq
\rho_n(\xi, N) = \sqrt{{2\over \phi}} {1\over 2^n n!} \sum_{j=-N/2}^{N/2}
e^{-2\pi (\xi-j)^2/\phi}\left( H_n \left(\sqrt{{2\pi\over
\phi}}(\xi-j)\right)\right)^2
\label{hdensity}
\eeq
where the flux $\phi$ and the particle number $N$ are related as in (\ref{deg})
for full filling.

\begin{figure}
\epsffile{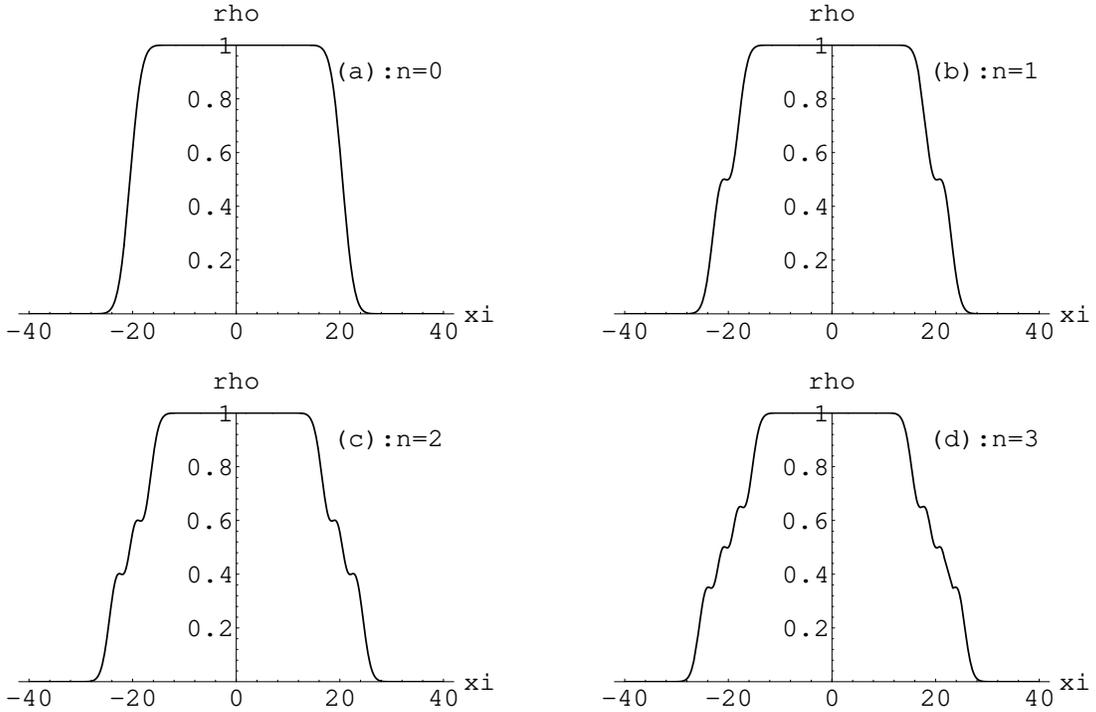}
\caption{The Landau gauge density functions (\protect{\ref{hdensity}}) plotted
for $N=40$ electrons for the first four Landau levels, $n=0,1,2,3$. Notice the
uniform bulk density equal to $1$ and the boundary fall-off at $\xi\simeq
\pm{N\over 2}$. For the $n^{th}$ Landau level there are $n$ distinct steps at
each boundary.}
\label{hermplot}
\end{figure}

\begin{figure}
\epsffile{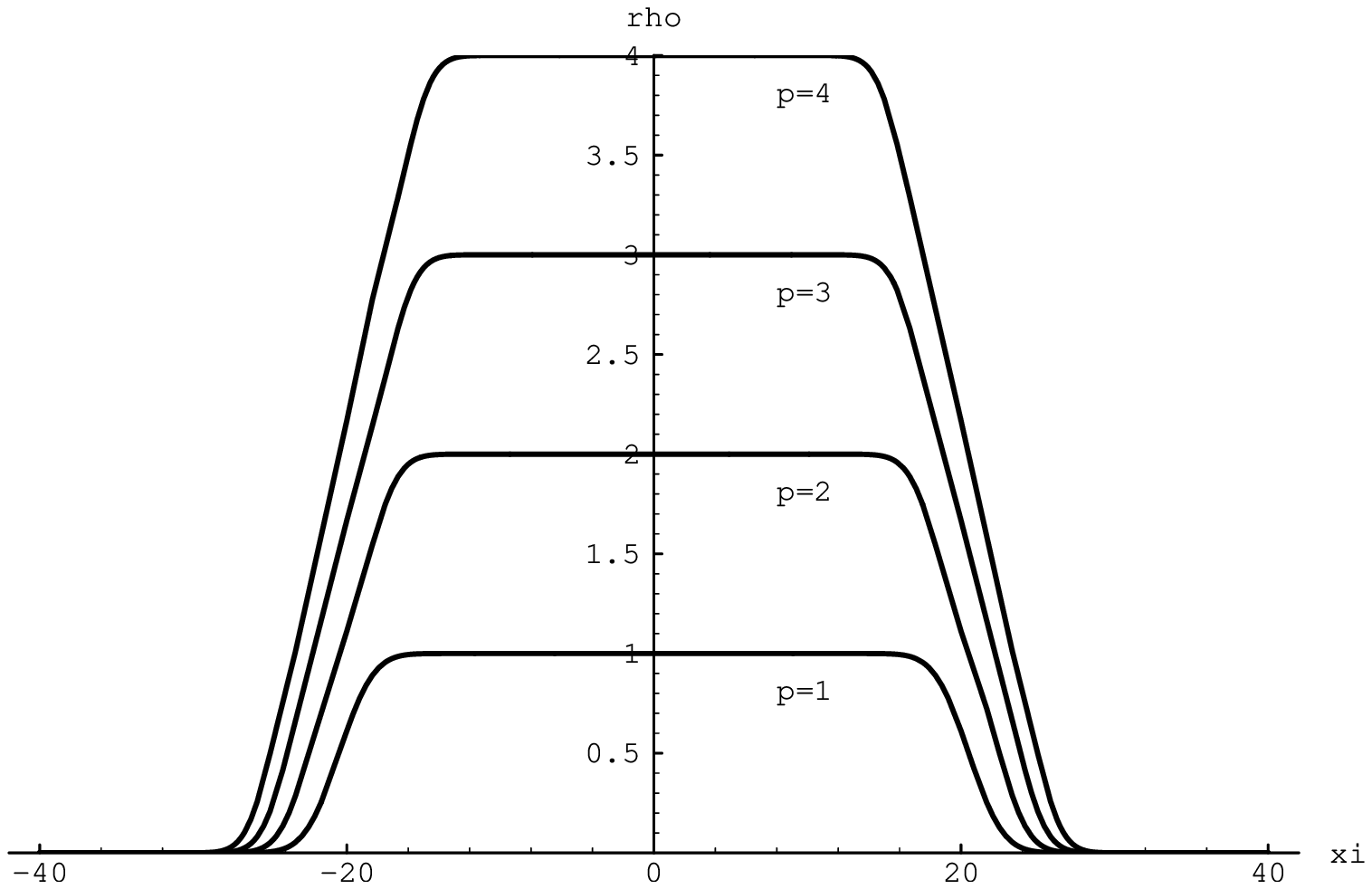}
\caption{The Landau gauge density functions $\rho_{\rm total}^{(p)}(\xi, N)$
for $p$ fully filled Landau levels, plotted for $N=40$ electrons. Notice the
uniform bulk density equal to $p$, the boundary density equal to ${p\over 2}$,
and the absence of the boundary steps seen in the individual higher Landau
level densities in Figure (\protect{\ref{hermplot}}).}
\label{hfullplot}
\end{figure}

The density (\ref{hdensity}) is plotted in Figure (\ref{hermplot}) for the
first four Landau levels for $N=40$ and $\phi \simeq N$.\footnote{Taking
$\phi=N-\mu$ with $\mu=O(1)$ does not affect these plots in any noticeable
manner. Also note that I have taken $N=40$ rather than $N=100$ simply due to a
limited graphics-plotting capacity.} From these plots we observe the same
behaviour as in the symmetric gauge. In the {\it lowest} Landau level, there is
a uniform bulk density equal to 1, which drops rapidly to zero in the vicinity
of the boundaries at $\xi =\pm{\phi\over 2} \simeq\pm {N\over 2}$. In the {\it
higher} Landau levels there is once again distinct step-like behaviour at the
edge - specifically, in the $n^{\rm th}$ Landau level there are $n$ distinct
steps in the density in the vicinity of the boundaries. Moreover, when the
densities for successive Landau levels are summed, as in (\ref{full}), these
boundary steps cancel out, leaving a smooth boundary density of the same form
as the lowest Landau level case: see Figure (\ref{hfullplot}).

It is much more difficult to prove rigorous results for the asymptotics of the
Landau gauge $\rho_n(\xi, N)$ because there are no clear analogues of the
Laguerre polynomial results ({\ref{der}},{\ref{induction}}) for the Hermite
polynomials. However, the Christoffel-Darboux identity for the Hermite
polynomials \cite{bat} implies the following {\it finite} sum formula:
\beq
\sum_{n=0}^{m} {(H_n(x))^2\over 2^n n!}={1\over 2^{m+1} m!}\left(
(H_{m+1}(x))^2-H_{m}(x) H_{m+2}(x)\right)
\label{christoffel}
\eeq
Applying this formula to the fully filled density $\rho_{\rm
total}^{(p)}\equiv\sum_{n=0}^{p-1}\rho_n$ with $\rho_n$ as in (\ref{hdensity})
we see immediately that
\beq
\rho_{\rm total}^{(p)}=p \rho_p + \sqrt{{2\over \phi}}{1\over 2^p (p-1)!}
\sum_{j=-N/2}^{N/2} e^{-2 \pi (\xi-j)^2/\phi} H_{p-1}\left(\sqrt{{2\pi\over
\phi}}(\xi-j)\right) H_{p+1}\left(\sqrt{{2\pi\over \phi}}(\xi-j)\right)
\label{darboux}
\eeq
The summation term on the right hand side of (\ref{darboux}) is concentrated on
the boundary - the fact that it vanishes in the bulk can be seen most directly
by noting that the orthogonality of the polynomials $H_{p-1}$ and $H_{p+1}$
ensures that the leading contribution to the Euler-MacLaurin formula \cite{cmb}
vanishes. Therefore, the bulk value of the fully filled density is
asymptotically equal to $p$ times the bulk density of the $p^{th}$ Landau
level. This bulk value (for any given Landau level) may be evaluated deep in
the bulk at $\xi=0$ by applying the Euler-MacLaurin formula to $\rho_n(0, N)$
\barr
\rho_n(0, N)&=&\sqrt{{2\over \phi}} {1\over 2^n n!} \sum_{j=-N/2}^{N/2}
e^{-2\pi j^2/\phi}\left( H_n \left(\sqrt{{2\pi\over
\phi}}j\right)\right)^2\nonumber\\
&=&\sqrt{{2\over \phi}}\left\{ {1\over 2^n n!}\int_{-N/2}^{+N/2} dx e^{-2\pi
x^2/\phi}\left( H_n \left(\sqrt{{2\pi\over \phi}}x\right)\right)^2
\right.\nonumber\\
& &\hspace{1in}\left.+ {1\over 2^n n!} e^{-2\pi (N/2)^2/\phi}\left( H_n
\left(\sqrt{{2\pi\over \phi}}{N\over 2}\right)\right)^2
+ \dots\right\}\nonumber\\
&=&1+O\left({1\over \sqrt{N}}\right)
\label{euler}
\earr
where we have used the fact that ${1\over \sqrt{2^n n!}}|e^{-x^2/2}H_n(x)|< k$
for all $x$ and $n$, where $k=1.086435...$ is Charlier's number \cite{bat}. By
a similar computation, now evaluating $\rho_n(\xi, N)$ at the edge where
$\xi\simeq\pm {N\over 2}$, we find
\barr
\rho_n(\pm {N\over 2}, N) &=& \sqrt{{2\over \phi}} {1\over 2^n n!}
\sum_{j=-N/2}^{N/2} e^{-2\pi (j \mp N/2)^2/\phi}\left( H_n
\left(\sqrt{{2\pi\over \phi}}(j\mp N/2)\right)\right)^2\nonumber\\
&=&{1\over 2} +O\left({1\over \sqrt{N}}\right)
\label{macl}
\earr
\b
\section{Conclusion}
\b
In this paper it has been shown that, in the $N\to\infty$ limit, the
expectation value of the planar electron density for the $n^{th}$ Landau level
exhibits $n$ distinct steps near the boundary. This has been shown in both the
symmetric gauge (appropriate for circular samples) and in the Landau gauge
(appropriate for rectangular samples), in order to dispel possible doubts that
this is perhaps an artifact of the particular (gauge dependent) basis used for
the degenerate subspaces. At first sight, such non-smooth boundary structure
seems to be problematic for the usual \cite{wen,sto,cdtz,iso} identification of
the boundary charge with a chiral Kac-Moody current. However, when {\it
successive} Landau levels are filled these boundary steps cancel out, leaving a
smooth boundary whose boundary characteristics have been computed exactly (see
equations (\ref{gen},\ref{ppslope})). It is not {\it a priori} obvious, in this
independent particle picture, that the densities for independent Landau levels
{\it must} conspire to produce a smooth edge behaviour.

Having identified the boundary behaviour of the higher Landau level electron
densities, it would now be interesting to consider the boundary currents within
a given Landau level, as discussed, for example, in \cite{mar} for the lowest
Landau level. Presumably, for successively filled Landau levels one obtains
different representations of the $W_{1+\infty}$ algebra. It would be very
interesting to learn whether there is any connection between the boundary steps
of the $n^{th}$ Landau level density and the different "boundaries" needed for
the $W_{1+\infty}$ representation theory and for the fractional QHE hierarchy
\cite{wen,zee,cap}. It should also be possible to apply some of the results
obtained here, for the asymptotic behaviour of sums of Laguerre and Hermite
polynomials, to the analysis of the pair correlation function \cite{mac} and to
the Coulomb interaction energy in higher Landau levels.

\b
\vskip 2in

\noindent{\bf Acknowledgements:} I am grateful to C.~Bender, A.~Cappelli,
A.~Lerda, C.~Trugenberger and G.~Zemba for discussions relating to this
problem. This work has been supported in part by the D.O.E. through grant
number DE-FG02-92ER40716.00, and by the University of Connecticut Research
Foundation.


\newpage
\clearpage


\end{document}